\documentclass{ifacconf}
\usepackage{graphicx}      
\usepackage{natbib}        
\usepackage{amssymb}
\usepackage{amsmath}
\usepackage{xspace}
\usepackage[table]{xcolor}
\usepackage{booktabs}
\usepackage{tikz}
\usepackage{enumitem}
\newlist{mylist_a}{enumerate}{1}
\setlist[mylist_a]{label=\textsc{a}\oldstylenums{\arabic*}}
\usepackage{listings}
\usepackage{textcomp}

\usetikzlibrary{arrows, positioning, fit, calc, backgrounds, trees}
\tikzset{>=latex}
\tikzset{
	punkt/.style={
		rectangle,
		rounded corners,
		draw=black, thick,
		text width=11.5em,
		minimum height=3em,
		text centered, inner sep = 5pt},
	mybox/.style={
		rectangle,
		rounded corners,
		draw=black, thick,
		text width=9em,
		minimum height=3em,
		text centered, inner sep = 5pt,
		fill = black!20},
	mynewbox/.style={
		rectangle,
		rounded corners,
		draw=black, thick,
		text width=9em,
		minimum height=5em,
		text centered, inner sep = 5pt,
		fill = black!20},
	myotherbox/.style={
		rectangle,
		rounded corners,
		draw=black, thick,
		dashed,
		text width=9em,
		minimum height=3em,
		text centered, inner sep = 5pt,
		fill = white},
	mysmallbox/.style={
		rectangle,
		rounded corners,
		draw=black, thick,
		text width=6.0em,
		minimum height=5.5em,
		text centered, inner sep = 5pt,
		fill = white},
	myPolyChaosBox/.style={
		rectangle,
		rounded corners,
		draw=black, thick,
		text width=6.0em,
		minimum height=3em,
		text centered, inner sep = 5pt,
		fill = white},
	boundingbox/.style = {
		rounded corners,
		draw = black,
		draw=gray!50,
		inner sep = 9pt,
		fill = black!3
	},
	pil/.style={
		->,
		thick,
		shorten <=2pt,
		shorten >=2pt,},
	treenode/.style = {
		anchor=west,
		font={\scriptsize}
	}
}

\newcommand{\polychaos}{\emph{PolyChaos.jl}\xspace}

\newcommand{\pce}{\textsc{pce}\xspace}
\newcommand{\captionOffset}{-6mm}
\newcommand{\ev}[1]{\mathbb{E}\left(#1\right)}
\newcommand{\std}[1]{\sigma\!\left(#1\right)}

\newtheorem{rema}{Remark}
\newtheorem{exmpl}{Example}

\newcommand{\samplespace}{\Omega}
\newcommand{\sigmaalgebra}{\mathfrak{F}}

\newcommand{\meas}{\mu}

\newcommand{\basisfun}{\phi}

\newcommand{\pceIndex}{k}
\newcommand{\pceMax}{\hat{k}}
\newcommand{\pceIndexSet}{\mathcal{K}}

\newcommand{\polycoeff}{a}

\newcommand{\Ltwospace}{L^2(\samplespace,\meas;\mathbb{R})}

\newcommand{\normalization}{\gamma}

\newcommand{\code}[1]{\textit{#1}}

\newcommand{\rv}[1]{\mathsf{#1}}

\begin{document}
\begin{frontmatter}

\title{PolyChaos.jl -- A Julia Package for Polynomial Chaos in Systems and Control }

\thanks[footnoteinfo]{This  work  was  supported  by  the  Helmholtz Association  under  the  joint initiative  ``Energy System 2050 -- A  Contribution  of  the  Research  Field Energy.''\\
This work has been done while TF was with the Institute for Automation and Applied Informatics,
   Karlsruhe Institute of Technology, Karlsruhe, Germany.}

\author[First]{Tillmann M\"uhlpfordt}
\author[First]{Frederik Zahn}
\author[First]{Veit Hagenmeyer}
\author[Second]{Timm Faulwasser}

\address[First]{Institute for Automation and Applied Informatics,
   Karlsruhe Institute of Technology, Karlsruhe, Germany (e-mail: {tillmann.muehlpfordt, frederik.zahn, veit.hagenmeyer}@kit.edu).}
\address[Second]{Department of Electrical Engineering and Information Technology, Technical University Dortmund, Dortmund, Germany (e-mail: timm.faulwasser@ieee.org)}

\begin{abstract}                
Polynomial chaos expansion (\pce) is an increasingly popular technique for uncertainty propagation and quantification in systems and control.
Based on the theory of Hilbert spaces and orthogonal polynomials, \pce allows for a unifying mathematical framework to study systems under arbitrary uncertainties of finite variance; we introduce this problem as a so-called mapping under uncertainty.
For practical \pce-based applications we require orthogonal polynomials relative to given probability densities, and their quadrature rules.
With \polychaos we provide a Julia software package that delivers the desired functionality: given a probability density function, \polychaos offers several numerical routines to construct the respective orthogonal polynomials, and the quadrature rules together with tensorized scalar products.
\polychaos is the first \pce-related software written in Julia, a scientific programming language that combines the readability of scripted languages with the speed of compiled languages.
We provide illustrating numerical examples that show both \pce and \polychaos in action.

\end{abstract}

\begin{keyword}
polynomial chaos expansion, uncertainties, stochastic optimal control, Julia
\end{keyword}

\end{frontmatter}

\section{Introduction}
George Box's celebrated assessment that ``all models are wrong, but some are useful,'' see \citep{Box1979ModelsAreWrong}, may be read as an allusion to the importance of uncertainties for mathematical models: it is not just that mathematical models may be wrong qualitatively---e.g. failing to account for nonlinear phenomena---but the mathematical surrogates may be wrong also quantitatively---e.g. not being able to assign a precise numerical value to a chemical reaction rate.
In the present paper we deal with the latter case: we are aware that uncertainties are present, and we are aiming for computational methods to account for them explicitly.

The traditional approach to dealing with these kinds of uncertainties is to sample and to simulate the system for each realization---the so-called Monte Carlo method.
There exists a myriad of sampling-based methods that differ mostly with respect to how samples are generated and how many of them are required to capture the statistics, see \citep{Xiu10book, LeMaitre10book, Sullivan15book}.
However, sampling-based methods scale too poorly for online optimization and control applications.
Polynomial chaos expansion (\pce) is a viable alternative to facilitate uncertainty propagation and uncertainty quantification.
Dating back to \citep{Wiener38} \pce is a Hilbert space technique that expands random variables in terms of polynomials that are orthogonal relative to the underlying probability measure.
\cite{Xiu02} extended \citeauthor{Wiener38}'s work to beyond the Gaussian measure.
The advantage of \pce for applications in systems and control is that it renders infinite-dimensional problems finite-dimensional.
This facilitates the simulation of ordinary differential equations under uncertainty, stochastic optimal control problems, or optimization problems under uncertainty, see for instance \citep{HOVER2006789, Kim2013PCE, Paulson2017aPC, Paulson2015StabilityMPC, Fagiano12a, Muehlpfordt16a, Bradford2019FeedbackNMPC, Muehlpfordt18segan, Muehlpfordt2019b}.

Whenever \pce is applied, however, we need to know the orthogonal bases for given probability densities, and we need to solve integrals efficiently.
This computational overhead calls for efficient, easy-to-use, and well-documented software.
With \polychaos, which is available open source~\citep{Muehlpfordt2019PolyChaos}, we deliver the first \pce-related software in the Julia programming language: given an arbitrary probability density function, \polychaos allows to construct the orthogonal bases via several routines from \cite{Gautschi04book}.
Furthermore, quadrature rules to solve integrals or tensorized scalar products of the basis functions are available easily.
Although there exist several software packages that provide \pce functionality, e.g. \code{UQLab} (Matlab, \citep{Marelli2014}), \code{Chaospy} (Python, \citep{Feinberg2015}), or \code{OpenTURNS} (Python, \citep{Baudin2017}), none previously existed in the Julia programming language, see \citep{Bezanson2017}.
Julia is dedicated to scientific computing, and it aims to combine the readability of scripted languages with the performance of compiled languages.

Three main parts make up the present paper:
Section~\ref{sec:ProblemFormulation} covers the theoretical framework:
Given an input random variable, and given a suitable mapping, what is the image random variable?
Undeniably, we provide but a glance at the theory of \pce; we refer to the rich literature for more details, e.g. \citep{Xiu10book, LeMaitre10book, Sullivan15book,STREIF20131}.
The gist of Section~\ref{sec:ProblemFormulation} is to show the \pce-practitioner that the applicability of the method hinges on two key items: the orthogonal bases must be known, and integrals must be solved efficiently.
Section~\ref{sec:PolyChaos} introduces the concepts permeating \polychaos.
We focus on the concepts and omit implementational details as the code is available open source, see \citep{Muehlpfordt2019PolyChaos}.
Finally, in Section~\ref{sec:NumericalExamples} we show how to apply \polychaos to three numerical problems.

\section{Problem formulation}
\label{sec:ProblemFormulation}
\subsection{Setting}
We study mappings under uncertainty: given some input random variable, we are interested in the image random variable stemming from a known mapping.
\begin{prob}[Mapping under uncertainty]
\label{prob:MappingUnderUncertainty}
Let $(\samplespace, \sigmaalgebra, \meas)$ be a probability space with sample space~$\samplespace$, sigma algebra~$\sigmaalgebra$, and an absolutely continuous, non-negative probability measure~$\meas$.
Also, let $\Ltwospace$ be the Hilbert space of all (equivalence classes of) real-valued random variables of finite variance.
Given the random variable $\rv{x} \in \Ltwospace$ find the random variable $\rv{y} \in \Ltwospace$ that is defined implicitly via
\begin{equation}
\label{eq:ModelEquation}
\rv{0} = F(\rv{y}, \rv{x}),
\end{equation}
where $F(\cdot)$ is a suitable implicit mapping.
\end{prob}

The mapping $F(\cdot)$ in Problem~\ref{prob:MappingUnderUncertainty} may stand for the solution to a (discretized) ordinary differential equation, for the solution to a discrete-time system, for a (system) of nonlinear algebraic equations, or for an argmin operator.

\begin{rema}[Several sources of uncertainty]~\\
	For sake of readability we restrict our presentation to single sources of uncertainty, hence univariate polynomial bases.
	For $m$ independent sources of uncertainty we can construct the $m$-variate orthogonal basis from the product of the $m$ respective univariate bases, see \citep{Xiu10book, Sullivan15book}.
\end{rema}

\begin{rema}[Several random variables]~\\
	\label{rema:SeveralRandomVariables}
	For sake of notation both Problem~\ref{prob:MappingUnderUncertainty} consider a single input uncertainty~$\rv{x}$ that is mapped to a single image random variable~$\rv{y}$.
	The extension to several uncertainties $\rv{x}_i \in \Ltwospace$ for $i \in \{1, \hdots, n_x\}$ and several image random variables $\rv{y}_i$ for $i \in \{ 1, \hdots, n_y \}$ would lead to substituting~\eqref{eq:ModelEquation} with
	$$\rv{0} = F(\rv{y}_1, \hdots, \rv{y}_{n_y}, \rv{x}_1, \hdots, \rv{x}_{n_x}).$$
\end{rema}

Problem~\ref{prob:MappingUnderUncertainty} is infinite-dimensional, hence intrinsically challenging.
A popular method to render Problem~\ref{prob:MappingUnderUncertainty} tractable is polynomial chaos expansion (\pce): a Hilbert space technique for random variables that is mathematically equivalent to Fourier series for periodic functions.

\subsection{Polynomial chaos expansion}

The \pce of the random variable $\rv{x} \in \Ltwospace$ is
\begin{equation}
\label{eq:PCE_Definition}
\rv{x} = \sum_{\pceIndex \in \pceIndexSet} x_{\pceIndex} \basisfun_{\pceIndex},
\end{equation}
with $\pceIndexSet \subseteq \mathbb{N}_0$, and where $\{ \basisfun_{\pceIndex} \}_{\pceIndex \in \mathbb{N}_0}$ is an ordered set of monic orthogonal polynomials that forms a complete orthogonal sequence in $\Ltwospace$, see \citep{Xiu02, Sullivan15book}.
The polynomials $\basisfun_{\pceIndex}$ satisfy
\begin{subequations}
	\begin{align}
	\basisfun_0 &= 1, \\
	\basisfun_{\pceIndex}(\tau) &= \tau^{\pceIndex} + \polycoeff_{\pceIndex - 1} \tau^{\pceIndex - 1} + \hdots + \polycoeff_0, && \forall \pceIndex \in \mathbb{N}, \\
	\label{eq:OrthogonalityCondition}
	\langle \basisfun_i, \basisfun_j \rangle &= \int_{\mathbb{R}} \basisfun_i(\tau) \basisfun_j(\tau) \mathrm{d} \meas (\tau) = \normalization_{i} \delta_{ij}, && \forall i,j \in \mathbb{N}_0,
	\end{align}
\end{subequations}
where $\normalization_i > 0$, and $\delta_{ij}$ is the Kronecker-delta.
The \pce from~\eqref{eq:PCE_Definition} is, generally speaking, exact whenever the index set~$\pceIndexSet$ is equal to~$\mathbb{N}_0$.
In case we truncate the \pce from~\eqref{eq:PCE_Definition} to finitely many terms $\pceIndexSet = \{ 0, \hdots, \pceMax \} $ there may be a truncation error---which is minimal with respect to the induced norm of the space $\Ltwospace$, cf. \citep{Xiu02, Sullivan15book}.

Given Problem~\ref{prob:MappingUnderUncertainty} and given the \pce~\eqref{eq:PCE_Definition} of $\rv{x}$, what are the \pce coefficients~$y_{\pceIndex}$ of~$\rv{y} = \sum_{\pceIndex \in \pceIndexSet} y_{\pceIndex} \basisfun_{\pceIndex}$ such that
\begin{equation}
\label{eq:ModelEquationsPCE}
\rv{0} = F \left( \sum_{\pceIndex \in \pceIndexSet} y_{\pceIndex} \basisfun_{\pceIndex}, \sum_{\pceIndex \in \pceIndexSet} x_{\pceIndex} \basisfun_{\pceIndex} \right)
\end{equation}
holds, and how can we compute them?
Galerkin projection is an option.
The idea of Galerkin projection is to project the \pce-overloaded model~\eqref{eq:ModelEquationsPCE} onto every basis element of~$\{ \basisfun_{\pceIndex} \}_{\pceIndex \in \pceIndexSet}$, see \citep{Sullivan15book}.
This leads to a set of deterministic equations in the form of integrals
\begin{subequations}
	\label{eq:GalerkinProjection}
	\begin{align}
	0 &= \Bigg\langle F\left(\sum_{\pceIndex \in \pceIndexSet} y_{\pceIndex} \basisfun_{\pceIndex}, \sum_{\pceIndex \in \pceIndexSet} x_{\pceIndex} \basisfun_{\pceIndex}\right), \basisfun_m \Bigg\rangle \\
	&= \int_{\samplespace} F\left(\sum_{\pceIndex \in \pceIndexSet} y_{\pceIndex} \basisfun_{\pceIndex}(\tau), \sum_{\pceIndex \in \pceIndexSet} x_{\pceIndex} \basisfun_{\pceIndex}(\tau) \right) \basisfun_m(\tau) \mathrm{d} \meas(\tau)
	\end{align}
\end{subequations}
for all $m \in \pceIndexSet$.
There are two main approaches to solve the integral in~\eqref{eq:GalerkinProjection}: non-intrusive and intrusive approaches.
Non-intrusive approaches tackle the integral by quadrature, Monte Carlo integration, or least-squares.
Intrusive approaches modify the expression~\eqref{eq:GalerkinProjection} to derive integrals that are either simpler to evaluate or that admit an exact Gauss quadrature.

\begin{rema}[Other advantages of \pce]
	\label{rema:AdvantagesofPCE}
	\pce offers other advantages than facilitating uncertainty propagation for Problem~\ref{prob:MappingUnderUncertainty}: it neither relies on sampling nor is it restricted to a specific family of distributions such as Gaussian distributions.
	Also, moments of random variables are functions of the \pce coefficients, see \citep{Xiu10book, Sullivan15book}.
\end{rema}

\subsection{Revised setting}
\label{sec:SettingRevised}
Let us revisit Problem~\ref{prob:MappingUnderUncertainty}---mappings under uncertainty---in light of \pce.
\begin{prob}[Mapping under uncertainty using \pce]
	\label{prob:MappingUnderUncertaintyUsingPCE}
	\mbox{}
	Consider the setup from Problem~\ref{prob:MappingUnderUncertainty}.
	Additionally, let the \pce of the random variable $\rv{x}$ be given by $\rv{x} = \sum_{\pceIndex \in \pceIndexSet} x_{\pceIndex} \basisfun_{\pceIndex}$ for a known orthogonal basis~$\{ \basisfun_\pceIndex \}_{\pceIndex \in \pceIndexSet}$ and a known probability measure~$\meas$.
	Then, find the \pce coefficients~$y_{\pceIndex}$ of $\rv{y} = \sum_{\pceIndex \in \pceIndexSet} y_{\pceIndex} \basisfun_{\pceIndex}$ such that~\eqref{eq:GalerkinProjection} holds.
\end{prob}

\begin{exmpl}[Van de Vusse reaction under uncertainty]
	\label{exmpl:VanDeVusseSetup}
	$\phantom{hello}$
	Consider a van de Vusse reaction with two uncertain reaction rates, see \citep{Paulson2015StabilityMPC, Scokaert1998ConstrainedLQR}.
	The dynamics of the concentrations can be modeled as
	\begin{subequations}
		\label{eq:uncVandevusse}
		\begin{align}
		\dot{\rv{c}}_A & = - \rv{c}_A u   -\rv{r}_1 \rv{c}_A - r_3 \rv{c}_A^2, && \rv{c}_A(0) = \rv{c}_{A, 0},\\
		\dot{\rv{c}}_B & = - \rv{c}_B u + \rv{r}_1 \rv{c}_A-\rv{r}_2 \rv{c}_B , && \rv{c}_B(0) = \rv{c}_{B, 0},
		\end{align}
	\end{subequations}
	with uncertain reaction rates $\rv{r}_1, \rv{r}_2$, a certain reaction rate $r_3$, and a fixed dilution rate $u$.
	In general, the initial conditions may be uncertain too.
	Using the notation of Problem~\ref{prob:MappingUnderUncertaintyUsingPCE} and Remark~\ref{rema:SeveralRandomVariables} we have $(\rv{x}_1, \rv{x}_2, \rv{x}_3, \rv{x}_4) = (\rv{r}_1, \rv{r}_2, \rv{c}_{A,0}, \rv{c}_{B,0})$ and $(\rv{y}_1, \rv{y}_2) = (\rv{c}_{A}, \rv{c}_{B})$.
	Inserting the \pce for all $\rv{x}_i$ with $i \in \{ 1, 2, 3, 4 \}$ and $\rv{y}_i$ with $i \in \{ 1, 2\}$, the Galerkin projection~\eqref{eq:GalerkinProjection} for the system~\eqref{eq:uncVandevusse} becomes\footnote{The observant reader noticed that, strictly speaking, we are leaving the setting from Problem~\ref{prob:MappingUnderUncertainty}, i.e. the realm of mere Hilbert spaces, with the setting from~\eqref{eq:uncVandevusse}.
		The rigorous introduction of random ordinary differential equations of the form~\eqref{eq:uncVandevusse} is beyond the scope of this paper.
		Our focus is on the Galerkin-based reformulation and the quantities introduced by \pce.\label{footnote:StochasticProcesses}}
\begin{subequations}
	\label{eq:VanDeVusse_GalerkinProjection}
	\footnotesize
	\begin{align}
	\dot{c}_{A, \pceIndex_1} &{=} - c_{A,\pceIndex_1}u {-} \sum_{\pceIndex_2, \pceIndex_3 \in \pceIndexSet} (r_{1,\pceIndex_2} c_{A, \pceIndex_3} {+} r_3 c_{A, \pceIndex_2} c_{A, \pceIndex_3}) \nu_{\pceIndex_1 \pceIndex_2 \pceIndex_3} \\
	\dot{c}_{B, \pceIndex_1} &{=} - c_{B,i} u +\,\, \sum_{\pceIndex_2, \pceIndex_3 \in \pceIndexSet} (r_{1,\pceIndex_2} c_{A, \pceIndex_3} - r_{2, \pceIndex_2} c_{B, \pceIndex_3}) \nu_{\pceIndex_1 \pceIndex_2 \pceIndex_3}
	\end{align}
	\normalsize
for all $\pceIndex_1 \in \pceIndexSet$, and
\begin{equation}
\nu_{\pceIndex_1 \pceIndex_2 \pceIndex_3} = \langle \basisfun_{\pceIndex_1} \basisfun_{\pceIndex_2}, \basisfun_{\pceIndex_3} \rangle/ \langle \basisfun_{\pceIndex_1}, \basisfun_{\pceIndex_1} \rangle.
\end{equation}

The Galerkin projection of the initial conditions leads to
\begin{equation}
c_{A, \pceIndex}(0) = c_{A, 0, \pceIndex}, ~ c_{B, \pceIndex}(0) = c_{B, 0, \pceIndex}
\end{equation}
\end{subequations}
for all $\pceIndex \in \pceIndexSet$.
The Galerkin-projected system~\eqref{eq:VanDeVusse_GalerkinProjection} is a system of ordinary differential equations in terms of the \pce coefficients of the concentrations.
As a consequence of applying \pce to~\eqref{eq:uncVandevusse} we need to know the basis polynomials~$\{ \basisfun_{\pceIndex} \}_{\pceIndex \in \pceIndexSet}$ and the numbers $\nu_{\pceIndex_1 \pceIndex_2 \pceIndex_3}$.
\end{exmpl}

So far, the significance of Problem~\ref{prob:MappingUnderUncertaintyUsingPCE} is more theoretical than practical.
To apply Problem~\ref{prob:MappingUnderUncertaintyUsingPCE} in practice we have to address the assumptions it hinges on, namely
\begin{mylist_a}
	\item \label{item:OrthogonalBasis} the orthogonal basis polynomials are known, and
	\item \label{item:ScalarProducts} the integrals from~\eqref{eq:GalerkinProjection}---respectively the scalars
	\begin{equation}
	\langle \basisfun_{\pceIndex_1} \cdots \basisfun_{\pceIndex_{n-1}}, \basisfun_{\pceIndex_n} \rangle
	\end{equation}
	for some $n \in \mathbb{N}$---can be computed efficiently.
\end{mylist_a}
For several well-known and often-employed uncertainties the Askey scheme provides orthogonal polynomials, see Table~\ref{tab:AskeyScheme}; accompanying quadrature rules to solve the integrals~\eqref{eq:GalerkinProjection} follow from the Golub-Welsch algorithm, see \citep{Golub69}.
In case of arbitrary probability densities we can utilize the Stieltjes procedure or the Lanczos procedure to construct the orthogonal polynomials---and then compute the quadrature rule according to \cite{Golub69}, or similar Gauss-like quadratures such as Gauss-Radau or Gauss-Lobatto, see \citep{Gautschi2002}.

\begin{table}[]
	\scriptsize
	\centering
	\caption{Askey scheme for ``classical'' distributions (\cite{Xiu02}).\label{tab:AskeyScheme}}
	\begin{tabular}{l   l  l  l}
		\toprule
		Type            & Support                    &  $ \basisfun_\pceIndex(\tau)$  & Polynomial basis \\ \midrule
		Beta            & $(0, 1)$                 &  $ \operatorname{P}_\pceIndex^{(\beta-1,\alpha-1)}(2\tau-1) $  & Jacobi           \\
		Gamma           & $(0, \infty)$            & $\operatorname{L}_\pceIndex^{(\alpha-1)}(\beta \tau)$    & Gen. Laguerre         \\
		Gaussian 		& $(-\infty, \infty)$      & $\operatorname{He}_\pceIndex(\tau)$   & Hermite          \\
		Uniform         & $[0, 1]$                 &  $ \operatorname{P}_\pceIndex^{(0,0)}(2\tau-1)$  & Legendre         \\ \bottomrule
	\end{tabular}
\end{table}

\subsection{Construction of orthogonal polynomials}
\label{sec:ConstructionOfOrthogonalPolynomials}
Before we explain numerical procedures we need to introduce a core theorem related to orthogonal polynomials.

\begin{thm}[Recurrence relation]
	\label{thm:ThreeTermRecurrenceRelation}
	Let $\basisfun_{\pceIndex}$ with $k \in \mathbb{N}_0$ be the monic orthogonal polynomials with respect to the measure $\meas$.
	Then,
	\begin{subequations}
		\begin{align}
		\label{eq:ThreeTermRelation}
		\basisfun_{\pceIndex + 1}(\tau)  &= (\tau - \alpha_{\pceIndex}) \basisfun_{\pceIndex}(\tau) - \beta_{\pceIndex} \basisfun_{\pceIndex - 1}(\tau), \quad\pceIndex \in \mathbb{N}_0,\\
		\label{eq:ThreeTermRelationInitial}
		\basisfun_{-1}(\tau)&= 0, \quad \basisfun_{0}(\tau) = 1 \\
		\alpha_k &= \frac{\langle \tau \basisfun_{\pceIndex}, \basisfun_{\pceIndex} \rangle}{\langle \basisfun_{\pceIndex}, \basisfun_{\pceIndex} \rangle},
		\quad \beta_k = \frac{\langle \basisfun_{\pceIndex}, \basisfun_{\pceIndex} \rangle}{\langle \basisfun_{\pceIndex - 1}, \basisfun_{\pceIndex - 1} \rangle}
		\end{align}
	\end{subequations}
\end{thm}
\emph{Proof:}	See \cite[1.3.1]{Gautschi04book}. \hfill $\square$

We hence identify orthogonal polynomials by their sequence of recurrence coefficients $\{ (\alpha_{\pceIndex}, \beta_{\pceIndex} \in ) \}_{\pceIndex \in \mathbb{N}_0}$---which we seek to compute for a given measure~$\meas$.
Moment-based methods are one possibility: based on the moments of the underlying measure, the recurrence coefficients can be computed.
This is the idea of~\cite{Paulson2017aPC}.
Unfortunately, this often leads to ill-conditioned problems, see \citep{Gautschi04book}.
The Stieltjes procedure and the Lanczos procedure are numerically stable alternatives.

\subsubsection{Stieltjes procedure}
Theorem~\ref{thm:ThreeTermRecurrenceRelation} suggests a simple iterative procedure to compute the recurrence coefficents $\alpha_{\pceIndex}, \beta_{\pceIndex}$, the so-called Stieltjes procedure:
For $\pceIndex = 0$ we define $\beta_0 = \int \mathrm{d} \meas (\tau) = 1$, from which we can compute $\alpha_0 = \langle \tau \rangle$, cf. \eqref{eq:ThreeTermRelationInitial}.
Knowing $(\alpha_0, \beta_0)$ we obtain $\basisfun_1$ from \eqref{eq:ThreeTermRelation}.
For $k = 1$ we compute $\beta_1 = \langle \basisfun_1, \basisfun_1 \rangle$, from which we get $\alpha_1 = \langle \tau \basisfun_1, \basisfun_1 \rangle / \langle \basisfun_1, \basisfun_1 \rangle$.
Knowing $(\alpha_1, \beta_1)$ we can construct $\basisfun_2$ from \eqref{eq:ThreeTermRelation}.
We can repeat this procedure until a desired degree~$k$ is reached.

The Stieltjes procedure is straightfoward to implement.
All occurring integrals can be solved efficiently using quadrature rules.
In case of numerical issues such as over- or underflow \cite{Gautschi04book} suggests to scale the weights and polynomials.

\subsubsection{Lanczos procedure}
The Lanczos algorithm allows to tri-diagonalize a given symmetric matrix~$A$, see~\citep{Golub83book}.
More specifically, a real symmetric matrix~$A$ allows the transformation~$Q^\top A Q = T$, where~$Q$ is orthogonal and~$T$ is symmetric and tridiagonal.
Given the matrix~$A$, Lanczos' algorithm produces the matrices~$Q$ and $T$.
In light of orthogonal polynomials, the Lanczos \emph{procedure} means to construct~$A$ such that the output of the Lanczos algorithm is the Jacobi matrix from which the recurrence coefficients can be read off.
\cite{Gautschi04book} shows how to construct the matrix~$A$ from the quadrature rule employed to solve the integrals.\footnote{For details on the Lanczos algorithm itself we refer to \citep{Golub83book} and/or \citep{Gragg1984}.}

\subsubsection{Multiple discretization}
Sometimes the underlying absolutely continuous probability measure $\meas$ allows to decompose integrals of some function~$f$ into $m \in \mathbb{N}$ parts according to
\begin{equation}
\label{eq:MultipleDiscretization}
\int_{[a, b]} f(\tau) \mathrm{d}\meas(\tau) = \sum_{i = 1}^{m} \int_{[a_i, b_i]} f_i(\tau) \mathrm{d} \meas_i(\tau),
\end{equation}
where $[a, b] = \cup_{i = 1}^{m} [a_i, b_i]$ with $-\infty \leq a < b \leq \infty$.
For instance, this may be the case for mixture models or densities defined on disjoint intervals.
Assuming we can apply either the Stieltjes procedure or the Lanczos procedure to the $m$ measures $\meas_i$, \cite{Gautschi04book} proposes---in the spirit of \emph{divide et impera}---a heuristic algorithm to construct the orthogonal polynomials relative to the measure $\meas$.

\subsection{Intermediate summary}
Given an input uncertainty in terms of a continuous random variable of finite variance we are interested in image random variables stemming from a known mapping, see Problem~\ref{prob:MappingUnderUncertainty}.
To apply the procedure in practice we require tools that compute orthogonal polynomials given a probability density, and that compute quadrature rules to solve the integrals.
The main contribution of the present paper is to introduce the Julia package \polychaos---built for this very purpose.

\section{PolyChaos.jl}
\label{sec:PolyChaos}
With \polychaos we deliver a software package in the Julia programming language that provides numerical routines for \pce-related computations, specifically addressing items \ref{item:OrthogonalBasis} and \ref{item:ScalarProducts} from Section~\ref{sec:SettingRevised}.

\subsection{Existing software}
Table~\ref{tab:SoftwareForPCE} lists existing software packages for \pce.
Except for \code{Chaospy} and \polychaos these packages are full-fledged libraries for uncertainty quantification; \pce comprises just one module of many, and it is used mostly for non-intrusive applications.
Amongst the software from Table~\ref{tab:SoftwareForPCE} \code{UQLab} and \code{Dakota} provide the richest functionality, each coming with a superb documentation.
While the core functions of \code{UQLab} are closed source, the scientific methods surrounding \pce are all available under the \textsc{bsd}\xspace 3-clause license.
Furthermore, \code{UQLab} provides methods for basis-adaptive \pce based on~\citep{Blatman10}.
\code{Dakota} is a mature framework: currently at version 6.0, version 3.0 beta, for instance, dates back to 2001.
The functionality of \code{MUQ} and \code{UQToolkit} is comparable; unfortunately they do not allow to compute orthogonal polynomials for arbitrary probability densities.
\code{OpenTURNS} is a full-fledged uncertainty quantification framework that comes with rich and mathematically detailed documentation.
The solely \pce-centered Python package \code{Chaospy} comes with the least restrictive \textsc{mit}-license whilst providing the core \pce functionality that includes the computation of orthogonal polynomials for arbitrary probability densities.

Table~\ref{tab:SoftwareForPCE} positions \polychaos in the landscape of software packages for \pce: its premise is to support arbitrary probability densities, for which it provides not just the Stieltjes but also the Lanczos procedure based on~\citep{Gautschi04book}.
In case the density can be composed as a sum of individual densities---as is common for (Gaussian) mixture models---\polychaos provides a specific method for multiple discretization based on \citep{Gautschi04book}.
Moreover, \polychaos provides several quadrature rules.

\begin{table*}
	\centering
	\caption{Existing software packages for polynomial chaos expansion.\label{tab:SoftwareForPCE}}
	\scriptsize
	\begin{tabular}{p{0.08\textwidth}p{0.10\textwidth}p{0.31\textwidth}m{0.10\textwidth}l}
		\toprule
		Name & Language & Features for polynomial chaos expansion & License & Reference \\
		\midrule
		\code{UQLab} & Matlab &%
		- Classic and arbitrary distributions \newline
		- Stieltjes procedure \newline
		- Gauss and sparse quadrature \newline
		- Basis-adaptive sparse \pce \newline
		- Least-angle regression
		& \textsc{bsd}\xspace 3-clause & \cite{Marelli2014} \\
		\code{Chaospy} & Python &%
		- Classic and arbitrary distributions\newline
		- Gram-Schmidt, Stieltjes procedure \newline
		- Gauss quadrature, Clenshaw-Curtis
		& \textsc{mit}\xspace & \cite{Feinberg2015} \\
		\code{OpenTURNS} & Python &%
		- Classic and arbitrary distributions\newline
		- Stieltjes procedure \newline
		- Gauss quadrature
		& \textsc{gnu lpgl}	& \cite{Baudin2017}\\
		\code{Dakota} & C++ &%
		- Classic and arbitrary distributions\newline
		- Stieltjes, Gram-Schmidt, Chebyshev \newline
		- Gauss and sparse quadrature \newline
		- Stochastic collocation & \textsc{gnu lpgl}	& \cite{DakotaSoftware}\\
		\code{MUQ} & C++, Python &%
		- Classic distributions \newline
		- Gauss quadrature
		& n/a & \cite{Conrad2013} \\
		\code{UQToolkit} & C++, Python &%
		- Classic distributions \newline
		- Gauss quadrature \newline
		& \textsc{gnu lgpl} & \cite{Debusschere2016} \\
		\midrule
		\polychaos &  Julia & %
		- Classic and arbitrary distributions \newline
		- Stieltjes and Lanczos procedure \newline
		- Multiple discretization \newline
		- Gauss quadrature,\,Fej\'er,\,Clenshaw-Curtis \newline
		- Tensorized scalar products
		&  \textsc{mit} & \cite{Muehlpfordt2019PolyChaos}\\
		\bottomrule
	\end{tabular}
\end{table*}

\subsection{Why Julia?}
Julia is a just-in-time compiled programming language for scientific computing, see~\citep{Bezanson2017}.
Julia solves the so-called two-language problem, the undesirable situation in which a programmer creates prototypes in one language (often based on easy-to-read scripts or notebooks, e.g. Matlab or Python), then having to switch to a different language (often compiled, e.g. C/C++) to achieve fast execution times.
Julia is a platform for both: rapid prototypes with intuitive code design can be morphed into type-specific, high-performance code.
The fact that types of values need not be declared explicitly is one reason why Julia solves the aforementioned two-language problem: users may never feel the need to declare types, yielding code reminiscent of scripted languages, yet users \emph{can} leverage the full power and expressiveness of Julia's type system to write cleaner and fast code.
With \polychaos we provide the first Julia package dedicated to orthogonal polynomials, quadrature rules, and \pce.

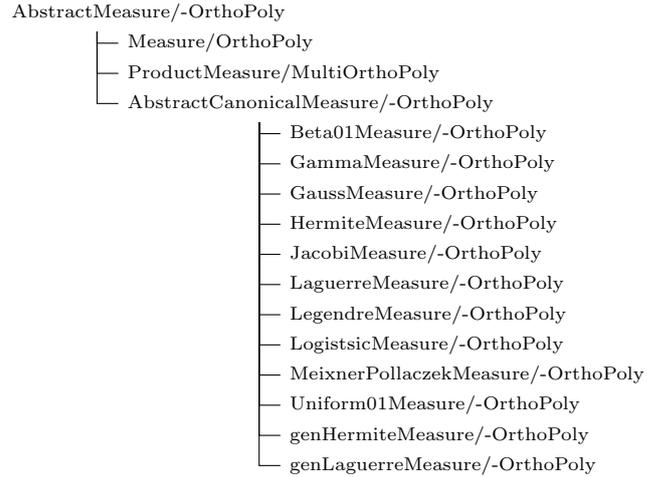
\begin{figure}
	\centering
	\begin{tikzpicture}[%
	grow via three points={one child at (-0.4,-0.4) and
		two children at (-0.4,-0.4) and (-0.4,-0.8)}, edge from parent path={(\tikzparentnode.200) |- (\tikzchildnode.west)}]
	\node[treenode] {AbstractMeasure/-OrthoPoly}
	child { node[treenode] {Measure/OrthoPoly}}
	child { node[treenode] {ProductMeasure/MultiOrthoPoly}}
	child { node[treenode] {AbstractCanonicalMeasure/-OrthoPoly}
		child { node[treenode] {Beta01Measure/-OrthoPoly}}
		child { node[treenode] {GammaMeasure/-OrthoPoly}}
		child { node[treenode] {GaussMeasure/-OrthoPoly}}
		child { node[treenode] {HermiteMeasure/-OrthoPoly}}
		child { node[treenode] {JacobiMeasure/-OrthoPoly}}
		child { node[treenode] {LaguerreMeasure/-OrthoPoly}}
		child { node[treenode] {LegendreMeasure/-OrthoPoly}}
		child { node[treenode] {LogistsicMeasure/-OrthoPoly}}
		child { node[treenode] {MeixnerPollaczekMeasure/-OrthoPoly}}
		child { node[treenode] {Uniform01Measure/-OrthoPoly}}
		child { node[treenode] {genHermiteMeasure/-OrthoPoly}}
		child { node[treenode] {genLaguerreMeasure/-OrthoPoly}}
	};
	\end{tikzpicture}
	\caption{Type hierarchy for measures and orth. polynomials.\label{fig:TypeHierarchyMeasureOrthoPoly}}
\end{figure}
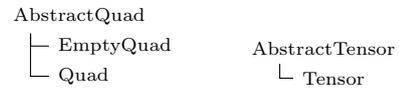
\begin{figure}
	\centering
	\begin{tikzpicture}[%
	grow via three points={one child at (-0.4,-0.4) and
		two children at (-0.4,-0.4) and (-0.4,-0.8)}, edge from parent path={(\tikzparentnode.200) |- (\tikzchildnode.west)}]
	\node[treenode] {AbstractQuad}
	child { node[treenode] {EmptyQuad}}
	child { node[treenode] {Quad}};
	\end{tikzpicture}
	\qquad
	\begin{tikzpicture}[%
	grow via three points={one child at (-0.4,-0.4) and
		two children at (-0.4,-0.4) and (-0.4,-0.8)}, edge from parent path={(\tikzparentnode.200) |- (\tikzchildnode.west)}]
	\node[treenode] {AbstractTensor}
	child { node[treenode] {Tensor}};
	\end{tikzpicture}
	\caption{Type hierarchy for quadrature rules and tensors.\label{fig:TypeHierarchyQuadTensor}}
\end{figure}

\subsection{Type hierarchy}
Every value in Julia has a type.
The conceptual foundation of the type system relies on \emph{abstract} types.
Abstract types serve but a single purpose: to form a type hierarchy.
It is neither desired nor possible to instantiate abstract types.
The type hierarchy remains independent from functions that operate on types.
To get what in other languages is called a struct or an object Julia provides \emph{composite} types.
A composite type has fields,\footnote{Methods can be fields too, the type of the field being \code{Function}.} it can be instantiated, and it can be declared a subtype of abstract types.

For \polychaos we devise our own type hierarchy.
Figure~\ref{fig:TypeHierarchyMeasureOrthoPoly} shows the two bread-and-butter type trees we need; abstract types carry the prefix ``Abstract.''
Take the abstract type \code{AbstractMeasure}: it has two composite subtypes: \code{Measure} and \code{ProductMeasure} with obvious meanings.
There exist, however, well-studied canonical measures such as Gaussian or uniform measures for which we introduce the abstract subtype \code{AbstractCanonicalMeasure}.
All subtypes of \code{AbstractCanonicalMeasure} are shown in Figure~\ref{fig:TypeHierarchyMeasureOrthoPoly}.
The type hierarchy for orthogonal polynomials mirrors that of measures: there are generic composite types for univariate polynomials, namely \code{OrthoPoly}, and multivariate polynomials, namely \code{MultiOrthoPoly}, and there are canonical orthogonal polynomials.
Figure~\ref{fig:TypeHierarchyQuadTensor} adds to the overall \polychaos type system quadrature rules via \code{AbstractQuad} and tensors of scalar products via \code{AbstractTensor}.
We emphasize once more that the type hierarchies from Figure~\ref{fig:TypeHierarchyMeasureOrthoPoly} and Figure~\ref{fig:TypeHierarchyQuadTensor} describe a \emph{concept} and no implementation.
The implementation details of all routines are beyond the scope of the present paper; we acknowledge that several routines are inspired by the Matlab code accompanying \cite{Gautschi04book}.

\section{Numerical examples}
\label{sec:NumericalExamples}
We consider three numerical examples that use the features of \polychaos.
The first example is about constructing the orthogonal basis for a non-trivial probability density;
the second example studies uncertainty propagation for Example~\ref{exmpl:VanDeVusseSetup};
the third example demonstrates optimal control for discrete-time systems under uncertainty.

The code for all numerical examples is available online.\footnote{See https://github.com/timueh/VanDeVusseUnderUncertainty\,.}

\subsection{Beta mixture -- Basis construction}
\label{sec:Example_BetaMixture}
\begin{figure}
	\centering
	\includegraphics[scale=1.0]{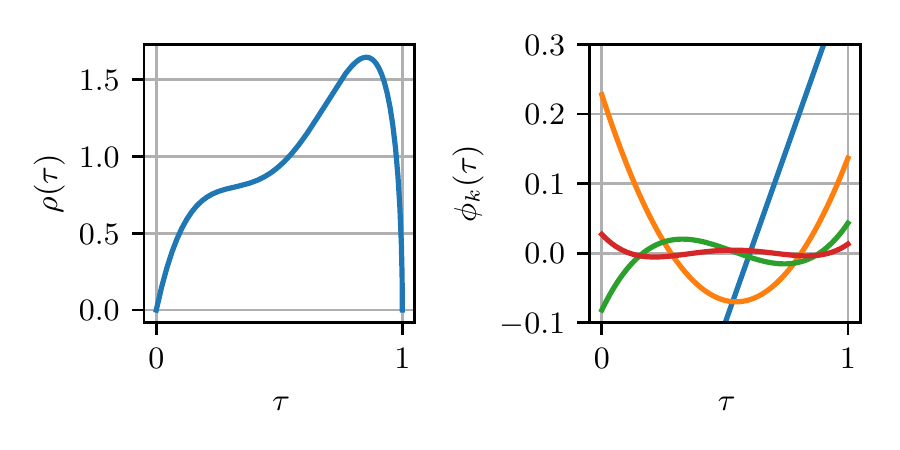}    
	\vspace{\captionOffset}
	\caption{Beta mixture density~$\rho(\tau)$ from~\eqref{eq:BetaMixtureDensity} and respective orthogonal polynomials $\basisfun_{\pceIndex}$ for degrees $\pceIndex \in \{1, 2, 3, 4\}$.\label{fig:BetaMixturePDF}}
\end{figure}
Consider a continuous random variable~$\rv{z} \in \Ltwospace$ with the absolutely continuous measure $\mathrm{d}\meas(\tau) = \rho(\tau) \mathrm{d}(\tau)$ being a Beta mixture with two components, hence the density for all $\tau \in (0, 1)$ is given by
\begin{equation}
\label{eq:BetaMixtureDensity}
\rho(\tau) = w_1 \rho_{B}(\tau; \alpha_1, \beta_1) + w_2 \rho_{B}(\tau; \alpha_2, \beta_2),
\end{equation}
with $w_1 + w_2 = 1$, and where $\rho_{B}(\tau; \alpha, \beta) = \tau^{\alpha - 1} (1 - \tau)^{\beta - 1} / B(\alpha, \beta)$ for all $\tau \in (0, 1)$ is a standard Beta density with positive shape parameters $\alpha, \beta$.
Figure~\ref{fig:BetaMixturePDF} shows the density~\eqref{eq:BetaMixtureDensity} for the specific values

\begin{center}
	\scriptsize
\begin{tabular}{cccccc}
	\toprule
	$w_1$ & $\alpha_1$ & $\beta_1$ 	& $w_2$ & $\alpha_2$ & $\beta_2$\\
	0.3 & 2.0 & 4.5  & 0.7 & 4.0 & 1.5 \\
	\bottomrule
\end{tabular}.
\end{center}
\normalsize

We employ multiple discretization from Section~\ref{sec:ConstructionOfOrthogonalPolynomials} to construct the first 5 polynomials that are orthogonal relative to~\eqref{eq:BetaMixtureDensity}.
This first requires to construct the orthogonal basis for each individual Beta distribution---for which we can use the built-in type \code{Beta01OrthoPoly}, see Figure~\ref{fig:TypeHierarchyMeasureOrthoPoly}.
Applying multiple discretization using the \polychaos function \code{mcdisretization()}, we obtain the following orthogonal polynomials for degrees 1 to 4
\begin{subequations}
	\label{eq:OrthogonalPolynomials}
	\begin{align}
	\basisfun_1(\tau) &= \tau - 0.6\\
	\basisfun_2(\tau) &= \tau^2 - 1.09 \tau + 0.23\\
	\basisfun_3(\tau) &= \tau^3 - 1.6 \tau^2 + 0.73 \tau - 0.08\\
	\basisfun_4(\tau) &= \tau^4 -2.11 \tau^3 + 1.47 \tau^2 - 0.38 \tau + 0.03,
	\end{align}
\end{subequations}
which are plotted in Figure~\ref{fig:BetaMixturePDF}.\footnote{Coefficients in \eqref{eq:OrthogonalPolynomials} are rounded to two decimals; Figure~\ref{fig:BetaMixturePDF} is based on all decimals.}

%

\subsection{Van de Vusse reaction -- Uncertainty propagation}
\label{sec:Example_Propagation}
\begin{figure}
	\centering
	\includegraphics[scale=.9]{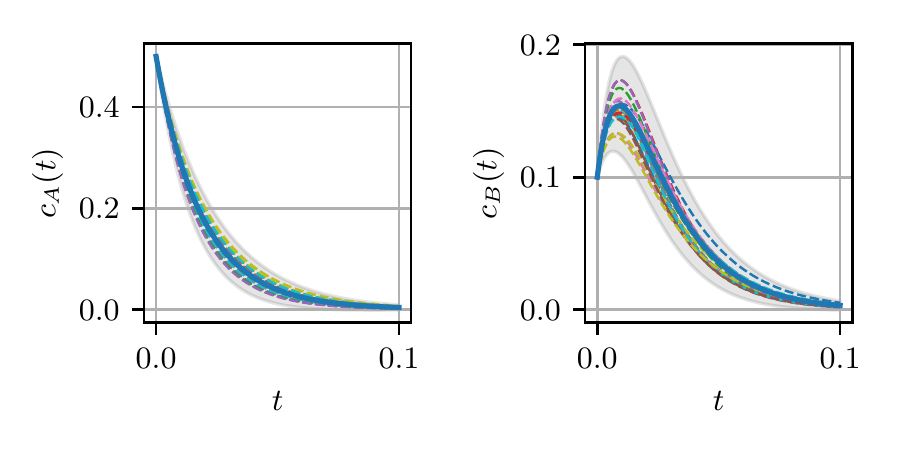}    
	\vspace{\captionOffset}
	\caption{Total of 20 realizations of the Van de Vusse reaction from Section~\ref{sec:Example_Propagation}; shaded area denotes the $\ev{\rv{c}_i(t)} \pm 3 \std{\rv{c}_i(t)}$ interval with $i \in \{A, B\}$.\label{fig:VanDeVusse_Propagation}}
\end{figure}

Recall Example~\ref{exmpl:VanDeVusseSetup} which results in the Galerkin-projected ordinary differential equations for the Van de Vusse reaction with uncertainties.
We choose the uncertain reaction ratees $\rv{r}_1$ and $\rv{r}_2$ each to follow an independent uniform distribution.
The parameters are
\begin{center}
	\begin{tabular}{ccccc}
		\toprule
		$\ev{\rv{r}_1}$ & $\std{\rv{r}_1}$ & $\ev{\rv{r}_2}$ & $\std{\rv{r}_2}$ & $r_3$\\
		$50$ & $0.1 \, \ev{\rv{r}_1}$ & $100$ & $0.1 \, \ev{\rv{r}_2}$ & $10$\\
		\bottomrule
	\end{tabular},
\end{center}
\normalsize
where the values for $\ev{\rv{r}_1}, \ev{\rv{r}_2}, r_3$ are taken from \cite{Scokaert1998ConstrainedLQR}.
For simplicity we choose deterministic initial conditions with
\begin{equation*}
\rv{c}_{A,0} \equiv c_{A,0} = 0.5, \quad
\rv{c}_{B,0} \equiv c_{B,0} = 0.1,
\end{equation*}
and a constant dilution rate of $u = 0.1$.
We construct the basis using the built-in type \code{Uniform01OrthoPoly} and compute the tensorized scalar products $\nu_{\pceIndex_1 \pceIndex_2 \pceIndex_3}$ from~\eqref{eq:VanDeVusse_GalerkinProjection} using the type \code{Tensor}, see Figures~\ref{fig:TypeHierarchyMeasureOrthoPoly} and~\ref{fig:TypeHierarchyQuadTensor}.
The set of differential equations~\eqref{eq:VanDeVusse_GalerkinProjection} is integrated using the Julia package \code{DifferentialEquations.jl}, see \citep{Rackauckas2017}.\footnote{In light of Footnote~\ref{footnote:StochasticProcesses} we remark that, generally speaking, we solve differential equations numerically by some form of discretization.
In that case we can view the task at hand as Problem~\ref{prob:MappingUnderUncertainty} together with Remark~\ref{rema:SeveralRandomVariables}.
This is true because random vectors can be viewed equivalently as discrete-time stochastic processes, see \cite{ Sullivan15book}.\label{footnote:SolvingODENumerically}}
From the solution of the \pce coefficients we can compute immediately moments, allowing to plot the $\ev{\rv{c}_i(t)} \pm 3 \std{\rv{c}_i(t)}$ interval with $i \in \{A, B\}$ without having to sample, cf. Remark~\ref{rema:AdvantagesofPCE}.
This is shown in Figure~\ref{fig:VanDeVusse_Propagation}, along with the solution trajectories for 20 realizations of the uncertainties that validate our findings.
The \pce-based mean $\ev{\rv{c}_i(t)}$ for $i \in \{A, B \}$ of the random-variable solutions corresponds to the solid line.
%
\begin{figure}
	\centering
	\includegraphics[scale=.9]{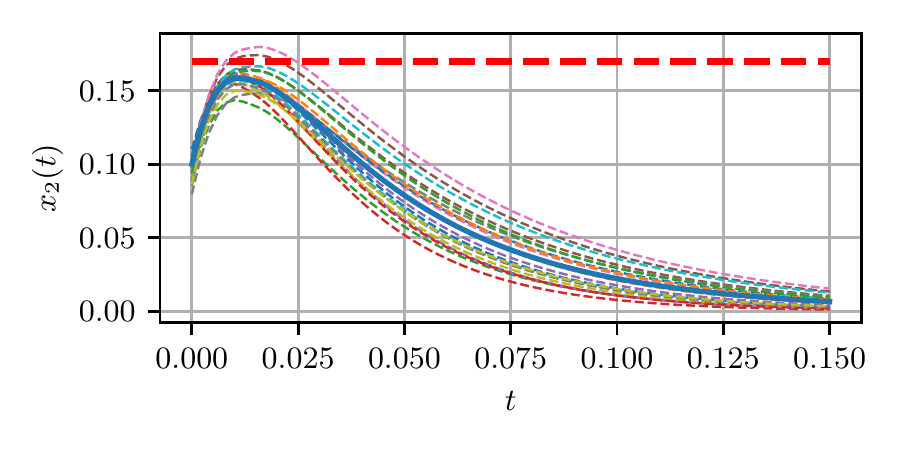}
	\vspace{\captionOffset}
	\caption{Total of 20 realizations of the second state of the linearized and discretized Van de Vusse reaction from Section~\ref{sec:Example_Optimization}.%
		\label{fig:VanDeVusse_Optimization}}
	\label{fig:optimization}
\end{figure}

\subsection{Van de Vusse reaction -- Stochastic optimal control}
\label{sec:Example_Optimization}
We consider the linearized and discretized van de Vusse reaction according to \cite{Paulson2015StabilityMPC}
\begin{subequations}
	\begin{equation}
	\label{eq:linearizedVandevusse}
	A = \begin{pmatrix}
	\rv{k} & 0\\
	0.088 & 0.819
	\end{pmatrix},
	~
	B = \begin{pmatrix}
	-0.005\\
	-0.002
	\end{pmatrix},
	\end{equation}
	where the parameter~$\rv{k}$ is uncertain according to
$
	\rv{k} = \underline{k} + (\overline{k} - \underline{k}) \rv{z},
$
\end{subequations}
where~$\rv{z}$ has the probability density from Section~\ref{sec:Example_BetaMixture}, and $(\underline{k}, \overline{k}) = (0.923, 0.926)$ is the support of~$\rv{k}$.\footnote{The support is chosen equivalent to the support from \cite{Paulson2015StabilityMPC}, where $\rv{k}$ is modeled as a single Beta distribution.}
The initial conditions $\rv{x}_{1,0}, \rv{x}_{2,0}$ of the system~\eqref{eq:linearizedVandevusse} are independent Gaussian random variables with
\begin{center}
	\begin{tabular}{cccc}
		\toprule
		$\ev{\rv{x}_{1,0}}$ & $\std{\rv{x}_{1,0}}$ & $\ev{\rv{x}_{2,0}}$ & $\std{\rv{x}_{2,0}}$ \\
		$1/2$ & $1/60$ & $1/10$ & $1/100$ \\
		\bottomrule
	\end{tabular}.
\end{center}
\normalsize
We wish to solve the following stochastic optimal control problem over the horizon of $\mathcal{T} = \{0, 1, \hdots, 74\}$
\begin{subequations}
	\label{eq:Example_Optimization}
	\begin{align}
		\min_{\substack{u(t) \\ \forall t \in \mathcal{T}}}& \sum_{t \in \mathcal{T}} \ev{\rv{x}(t+1)^{\top}Q\rv{x}(t+1)}+u(t)^{\top} R u(t)\\
		\nonumber
		\text{s.t.}&\\
		&\rv{x}(t + 1) = A \rv{x}(t)+ B u(t),  \quad \forall t \in \mathcal{T},\\
		&\rv{x}(0) = \begin{pmatrix}
			\rv{x}_{1,0}\\
			\rv{x}_{2,0}
			\end{pmatrix},	\\
		\label{eq:ReformulatedChanceConstraint}
		&\ev{\rv{\rv{x}_2(t)}} + \lambda \std{\rv{x}_2(t)} \leq \overline{x}_2, \quad \forall t \in \mathcal{T} \cup \{ 75 \},
	\end{align}
\end{subequations}
with positive definite weights $Q = I_2, R = 1$.\footnote{In light of Footnotes~\ref{footnote:StochasticProcesses} and~\ref{footnote:SolvingODENumerically} we remark that the random-variable discrete-time system from \eqref{eq:Example_Optimization} may be viewed either as a random vector or as a discrete-time stochastic process.}
The inequality constraint~\eqref{eq:ReformulatedChanceConstraint} is a reformulated chance constraint for $\lambda = 1.618$ and the upper limit $\overline{x}_2 = 0.17$, see \citep{Paulson2015StabilityMPC}.
We construct the basis using the built-in type \code{GaussOrthoPoly} twice together with the numerically computed basis from Section~\ref{sec:Example_BetaMixture}.
Galerkin projection can be applied to Problem~\eqref{eq:Example_Optimization}, which again leads to tensorized scalar products \citep{Paulson2015StabilityMPC}
\begin{equation}
\label{eq:PCEIngredients}
\langle \basisfun_{\pceIndex_1} \basisfun_{\pceIndex_2}, \basisfun_{\pceIndex_3} \rangle,
\langle \basisfun_{\pceIndex_1}, \basisfun_{\pceIndex_2} \rangle
\end{equation}
for all $\pceIndex_1, \pceIndex_2, \pceIndex_3 \in \pceIndexSet$, for which we use the type \code{Tensor}.
Figure~\ref{fig:VanDeVusse_Optimization} shows the trajectories of the second state for a total of 20 realizations of the uncertainties.
In Figure~\ref{fig:VanDeVusse_Optimization} the solid line denotes the \pce-obtained trajectory of the mean.
Owing to the chance constraint reformulation~\eqref{eq:ReformulatedChanceConstraint}, the bound~$\overline{x}_2 = 0.17$ may be violated.
For a total of 100,000 realizations (not shown) we empirically find that about 6\,\% of the trajectories violate the constraint.

\subsection{Computation times}
So far we showed the numerical results of three different examples.
We commented how to obtain the \pce-related quantities using \polychaos.
To assess the performance of \polychaos within each example we measured the times to construct the basis, and to compute the scalars~\eqref{eq:PCEIngredients} for 10,000 consecutive runs, including the first run to compile the code; Table~\ref{tab:ComputationTimes} shows the mean times obtained on a desktop computer with an Intel\textsuperscript{\textregistered} Core\texttrademark\xspace i7-8700 CPU 3.20GHz processor and 31\,GiB of RAM.
The maximum degree of the basis polynomials is 4 for all cases.
As we can see from Table~\ref{tab:ComputationTimes} the computation time overhead stemming from \polychaos is negligible for the considered examples.

\begin{table}
	\centering
	\caption{Computation times with \polychaos; $N_{\text{unc}}$ is the number of uncertainties.\label{tab:ComputationTimes}}
\scriptsize
    \begin{tabular}{lrrrr}
	\toprule
	& & \multicolumn{3}{c}{Mean time in $\mu$s for 10,000 runs} \\
	Example & $N_{\text{unc}}$ & Basis & $\langle \basisfun_{\pceIndex_1}, \basisfun_{\pceIndex_2} \rangle$ & $\langle \basisfun_{\pceIndex_1} \basisfun_{\pceIndex_2}, \basisfun_{\pceIndex_3} \rangle$  \\
	\midrule
	Beta mixture & 1 & 120  & 20    &    100   \\
	Propagation & 2 & 47       &     131    &  1,089  \\
	Optimization & 3 &      171    &  707    &  12,888 \\
	\bottomrule
	\end{tabular}
\end{table}

\section{Summary and outlook}
We introduce mappings under uncertainty and demonstrate how to solve them using polynomial chaos expansion, a Hilbert space technique for random variables of finite variance.
Polynomial chaos imposes mainly two computational burdens: finding orthogonal polynomials given a probability density, and determining their quadrature rules.
To facilitate these computations we introduce \polychaos, a software package written in the Julia programming language.
Three numerical examples demonstrate how to use \polychaos for specific mappings under uncertainty.
The computation time overhead stemming from \polychaos is negligible for the studied examples.
Future work will focus on comparing the computational speed and accuracy to other existing software packages.
Also, additional functionalities such as stochastic collocation and/or sparse basis construction are desirable to add.

{\footnotesize\bibliography{lit_cleaned_up}}

\end{document}